\documentclass[prd,aps,a4paper,nofootinbib,showpacs,twocolumn]{revtex4-1} 

\newif\ifusesec
\usesectrue  
   
\usepackage{graphicx} 
\usepackage{mathrsfs}
\usepackage{latexsym,amsmath,amsfonts,amssymb}
\usepackage{multirow}

\newcommand{\be}{\begin{equation}}
\newcommand{\ee}{\end{equation}}
\newcommand{\bea}{\begin{eqnarray}}
\newcommand{\eea}{\end{eqnarray}}

\newcommand{\D}{\partial}

\begin{document}

\title{Gravitational scattering, post-Minkowskian approximation \\ and Effective One-Body theory}

\author{Thibault \surname{Damour}}

\affiliation{Institut des Hautes Etudes Scientifiques, 91440 Bures-sur-Yvette, France}
\date{\today}

\begin{abstract}
A novel approach to the Effective One-Body description of gravitationally interacting two-body systems is introduced.
This approach is based on the {\it post-Minkowskian} approximation scheme 
(perturbation theory in $G$, without assuming small velocities), and employs a new dictionary focussing on the functional
dependence of the scattering angle on the total energy and the total angular momentum of the system. Using this approach,
we prove {\it to all orders in $v/c$}  two results that were previously  known to hold only to
a limited post-Newtonian accuracy: (i) the relativistic gravitational dynamics of a two-body system 
is equivalent, at first post-Minkowskian order, to the relativistic dynamics of an effective test particle moving in a Schwarzschild metric; and 
(ii) this equivalence requires the existence of an {\it exactly quadratic map} between the real 
(relativistic) two-body energy  and the (relativistic) energy of the effective particle. The same energy map is
also shown to apply to the effective one-body description of two masses interacting via tensor-scalar gravity.
\end{abstract}

\maketitle

\section{Introduction}

The Effective One-Body (EOB) formalism was conceived \cite{Buonanno:1998gg,Buonanno:2000ef,Damour:2000we,Damour:2001tu}
with the aim of analytically describing both the last few orbits of, and the complete gravitational-wave signal emitted by, coalescing
binary black holes. 
The predictions, made as early as 2000 \cite{Buonanno:2000ef}, by the EOB formalism have been broadly confirmed by subsequent numerical simulations \cite{Pretorius:2005gq,Campanelli:2005dd,Baker:2005vv,Buonanno:2006ui}. This then led to the development of numerical-relativity-improved versions
of the EOB dynamics and waveform (see, e.g., \cite{Damour:2007xr,Buonanno:2007pf,Damour:2007vq,Buonanno:2009qa,Damour:2012ky,Taracchini:2013rva,Damour:2014sva}),
which have helped the recent discovery, interpretation and data analysis of the first gravitational-wave signals
by the Laser Interferometer Gravitational-Wave Observatory \cite{Abbott:2016blz,TheLIGOScientific:2016qqj,Abbott:2016nmj}.

The aim of  the present paper is to introduce a novel theoretical approach to some of the basic structures of  EOB theory.
The hope is that this new approach could lead to theoretically improved versions of the EOB conservative dynamics, 
which might be useful in the upcoming era of high signal-to-noise-ratio gravitational-wave
observations. In this work we shall only consider the interaction of non-spinning bodies
at the first order in $G$. Our strategy is, however, generalizable to higher orders in $G$, and to spinning bodies.

The EOB conservative dynamics is a relativistic generalization of the well-known Newtonian-mechanics
fact that the relative dynamics of a two-body system (with masses $m_1$ and $m_2$) is equivalent to the motion of an effective particle of mass
\be
\mu \equiv \frac{m_1 m_2}{m_1+m_2} \, ,
\ee 
 submitted to the original two-body potential $V(| {\bf R}_1 -  {\bf R}_2|)$. In the case of the Newtonian gravitational
 interaction (i.e.   $V(| {\bf R}_1 -  {\bf R}_2|) = - G m_1 m_2/| {\bf R}_1 -  {\bf R}_2|$), the identity
 $m_1 m_2 = \mu (m_1+m_2)$ implies that the effective particle of mass $\mu$ moves in the gravitational potential of a  central mass
 equal to
 \be
 M \equiv m_1 +m_2 \, .
 \ee

The historical approach to defining the EOB (conservative) dynamics
\cite{Buonanno:1998gg,Damour:2000we,Damour:2001tu} has been, so far, based on two basic ingredients:

 (i) a post-Newtonian (PN) description of the two-body dynamics (whose limit when $c \to \infty$ is the Newtonian result just recalled); and

(ii)   a dictionary between the PN-expanded knowledge
of the two-body {\it bound states} (ellipticlike motions), and the bound states of a test particle moving in some external
metric. 

Here, these two historical ingredients will be replaced by two other ones. 

\begin{enumerate}

\item The PN approximation method
(which is a combined expansion in $\frac{G}{c^2}$  and in $\frac{1}{c^2}$) will be replaced by the {\it post-Minkowskian} (PM)
approximation method, i.e. by an expansion in the gravitational constant $G$, which never assumes that the
velocities are small compared to the velocity of light $c$. After some pioneering work in the late 1950's \
\cite{Bertotti:1956,BertottiPlebanski:1960},
the PM approach to gravitational motion has played a useful role around
the 1980's in clarifying the computation of retarded interactions (and associated radiation reaction) in binary systems
\cite{Rosenblum:1978zr,Westpfahl:1979gu,Bel:1981,Damour:1981bh}.

\item The bound states of two-body systems will be replaced by {\it scattering states}. The replacement of bound states by scattering 
states will oblige us to replace the usual dictionary of EOB theory by a new one (that we shall prove to be physically equivalent).
This will allow us to exploit the fully relativistic results on gravitational scattering motions 
\cite{Westpfahl:1979gu,Bel:1981,Portilla:1979xx,Portilla:1980uz,Westpfahl:1985,Westpfahl:1987,Ledvinka:2008tk} 
obtained by PM calculations.

 \end{enumerate}
 
 Before expounding our new strategy we shall briefly recall some of the basic features of the current approach to EOB theory.

\section{Reminder of the EOB dictionary for  bound states} \label{sec2}

The construction of the EOB  dynamics \cite{Buonanno:1998gg,Buonanno:2000ef,Damour:2000we,Damour:2001tu,Damour:2015isa} has been so far based on a  dictionary between the bound states (ellipticlike motions) of a gravitationally interacting two-body system, considered in the 
center of mass (cm) frame, and the bound states of an effective particle moving in an effective metric $g_{\mu \nu}^{\rm eff}$. Inspired by the Bohr-Sommerfeld
quantization conditions of bound states ($I_i = n_i \hbar$), the latter dictionary requires the identification between the action integrals 
$I_i = \frac1{2\pi} \oint P_i d Q^i$ (no sum on $i$) of the real and effective dynamics, i.e. (considering, for simplicity, the reduction
of the dynamics to the plane of the motion)
\be \label{actiondictionary}
I_R^{\rm eff} = I_R^{\rm real} \ ; \ I_{\varphi}^{\rm eff} = I_{\varphi}^{\rm real} \, .
\ee
On the other hand, EOB theory allows for an arbitrary (a priori undetermined) {\it energy map} $f$ between the real cm energy
${\mathcal E}_{\rm real}$ and the effective one ${\mathcal E}_{\rm eff}$:
\be \label{energymap}
{\mathcal E}_{\rm real} \ \stackrel{f}{\to} \ {\mathcal E}_{\rm eff} \, .
\ee
When dealing with bound states (which can be treated within PN theory, i.e. at successive orders in an expansion in $\frac1{c^2}$),
one parametrizes the unknown energy map $f$ by a PN expansion of the type
\bea \label{fexp}
\frac{{\mathcal E}_{\rm eff}}{m_0 c^2}&=& 1+ \frac{E_{\rm real}}{\mu c^2} \left( 1 + \alpha_1 \frac{E_{\rm real}}{\mu c^2} + 
 \alpha_2 \left( \frac{E_{\rm real}}{\mu c^2}\right)^2 \right. \nonumber \\
 &+& \left. \alpha_3 \left( \frac{E_{\rm real}}{\mu c^2}\right)^3 + \alpha_4 \left( \frac{E_{\rm real}}{\mu c^2}\right)^4 + \cdots  \right) \,,
\eea
where $m_0$ denotes the mass of the effective particle, and where $E_{\rm real}$ denotes the ``non-relativistic" real energy, i.e. the difference 
\be
E_{\rm real} \equiv {\mathcal E}_{\rm real} - (m_1 + m_2)c^2 \,.
\ee

In EOB theory,  the energy map $f$ is  determined, {\it at any given PN approximation}, by the basic EOB bound-state requirement
\be \label{basicEOBbound}
{\mathcal E}_{\rm eff}(I_R,  I_{\varphi})= f\left[ {\mathcal E}_{\rm real}(I_R,  I_{\varphi}) \right] \,,
\ee
in which both energies are expressed in terms of the {\it common} values of the action variables $I_i = I_i^{\rm eff} = I_i^{\rm real}$.

The so-determined energy map turns out to exhibit, so far, a very simple (and natural \cite{Itzykson:1980rh,Damour:1997ub}) structure, described,
at the presently known order,
by the following coefficients in the PN expansion \eqref{fexp}
\be
 \alpha_1= \frac{\nu}{2} \, ; \, 0= \alpha_2=\alpha_3= \alpha_4\, ,
\ee
where $\nu$ denotes the symmetric mass ratio,
 \be
 \nu  \equiv \frac{\mu}{M} = \frac{m_1 m_2}{(m_1+m_2)^2} \, .
 \ee 
More precisely,  the first two results  $\alpha_1= \frac{\nu}{2} \, ; \, \alpha_2=0$, which refer to the first post-Newtonian (1PN) and 
second post-Newtonian (2PN) levels, respectively,
were derived in Ref. \cite{Buonanno:1998gg}. At the 3PN level, and under the assumption already made in 
\cite{Buonanno:1998gg} that the effective
metric coincides,  at linear order in Newton's constant $G$, with the Schwarzschild metric, Ref. \cite{Damour:2000we} 
derived the next term in the energy map, and found it simply to be $\alpha_3=0$. Recently, the extension of the EOB
formalism to the 4PN level \cite{Damour:2015isa}, again found an uncorrected energy map, i.e. $\alpha_4=0$.
These results suggest, without, however,  proving it, that the energy map $f$ will remain uncorrected by higher-order PN
corrections, and is exactly quadratic.

One of the main motivations for the present extension of EOB theory to scattering states (i.e. hyperboliclike motions)
is to bypass the need to determine the energy map in the form of the PN expansion \eqref{fexp}. This will allow us to prove,
for the first time, that all the higher-order PN coefficients $\alpha_n$  in the PN expansion \eqref{fexp} actually vanish,
so that the energy map ${\mathcal E}_{\rm eff} = f({\mathcal E}_{\rm real})$ is exactly quadratic.
We shall also prove what was before assumed, namely that 
 the effective metric must coincide,  at linear order in $G$, with the Schwarzschild metric. [Ref. \cite{Buonanno:1998gg}
 explicitly assumed that the coefficient $b_1$ parametrizing the spatial part of the effective metric to order $G^1$ was equal
 to its Schwarzschild counterpart. The attempt of Ref. \cite{Damour:2000we} (see Appendix A there) to relax this assumption
 did not lead to a convincing result, so that Ref. \cite{Damour:2000we} finally kept this assumption on $b_1$.]
 
\section{Dictionary for scattering states} \label{sec3}

When considering scattering states (i.e. hyperboliclike motions) we lose the possibility of uniquely parametrizing two-body bound states
by means of the two action variables $I_R,  I_{\varphi}$. Actually, we still can use the second action variable, because
\be
I_{\varphi} \equiv \frac1{2\pi} \oint P_{\varphi} d {\varphi} = P_{\varphi} \equiv J\,,
\ee
is simply the total (cm) angular momentum of the binary system (considered as moving in the equatorial plane $\theta= \frac{\pi}{2}$). [As we restrict here our attention to non-spinning systems,
we could indifferently denote $P_{\varphi}$, which is the total {\it orbital} angular momentum, by $J$ or $L$.] We then
evidently keep the second half of the bound-state dictionary \eqref{actiondictionary}, namely
\be \label{Jdic}
J_{\rm eff} = J_{\rm real} =J
\ee
within our new scattering-state dictionary.

We need, however, a replacement for the first, radial half of the bound-state dictionary \eqref{actiondictionary}. This replacement is easily
found if we recall that the radial action variable $I_R$ is linked to the  evolution of the polar angle ${\varphi}$ within the plane of the motion. Indeed, Hamilton-Jacobi theory tells us to differentiate the energy-reduced action 
$S_0(R, {\varphi}; {\mathcal E}, J) = J \varphi + \int dR P_R(R; {\mathcal E}, J)$ with respect to $J$ to obtain the functional link between $\varphi$  and $R$,
namely
\be
\varphi = - \int dR \frac{\partial P_R(R ; {\mathcal E}, J)}{\partial J} + {\rm const.} \,.
\ee
In the bound-state case, the latter equation (which is gauge-dependent) yields, upon integration over a
radial period, the gauge-independent link
\be \label{Phibound}
\Phi_{\rm bound}=  -\oint dR \frac{\partial P_R(R ; {\mathcal E}, J)}{\partial J} = - 2 \pi \, \frac{\partial I_R({\mathcal E}, J)}{\partial J} \,,
\ee
between the total periastron precession per orbit, $\Phi_{\rm bound}$, and the $J$-derivative of the radial action, $I_R=\frac1{2\pi}\oint dR \,P_R$.

In the scattering case, it is very natural to replace the consideration of $\Phi_{\rm bound}$, Eq. \eqref{Phibound}, by that of the total angular change during scattering, i.e.
\be
\Phi_{\rm scatt}=  -\int_{- \infty}^{+\infty} dR \frac{\partial P_R(R; {\mathcal E}, J)}{\partial J} \,,
\ee
where the label ${- \infty}$ refers to the incoming state (at ${- \infty}$ in time, and ${+ \infty}$ in $R$), while the label ${+ \infty}$
refers to the outgoing state (at ${+ \infty}$ in time, and ${+ \infty}$ in $R$). The total change $\Phi_{\rm scatt}$ in polar angle
is usually parametrized in terms of the corresponding ``scattering angle" $\chi$ defined simply as
\be
\chi \equiv \Phi_{\rm scatt} - \pi \, ,
\ee
so that it vanishes for free motions.

In view of the links we just recalled, it is very natural, when considering scattering states, to replace the
 bound-state dictionary \eqref{actiondictionary} by the two conditions: Eq. \eqref{Jdic} together with
 \be \label{chi=}
 \chi_{\rm eff} = \chi_{\rm real}\,.
 \ee
 This leads us to replace the basic bound-state requirement \eqref{basicEOBbound}, by the
 following scattering-state requirement
 \be \label{dic1}
{\mathcal E}_{\rm eff}(\chi,  J)= f\left[ {\mathcal E}_{\rm real}(\chi, J) \right]\,,
\ee
or, equivalently,
 \be \label{dic2}
\chi_{\rm eff}({\mathcal E}_{\rm eff}, J)= \chi_{\rm real} ({\mathcal E}_{\rm real}, J) \,,
\ee
where it is understood that ${\mathcal E}_{\rm eff}$ (on the left-hand side) and ${\mathcal E}_{\rm real}$ (on the right-hand side) must be related by
the (looked-for, exact) energy map $f$, i.e.
\be
{\mathcal E}_{\rm eff}= f\left(  {\mathcal E}_{\rm real}   \right)\,.
\ee
Some comments on this scattering dictionary are in order. First, let us note that $\chi_{\rm real}$ measures the 
gravitational two-body scattering {\it in the cm frame}: it is the common value of the scattering angles
of each particle in the latter frame (where  ${\mathbf P}_1=-{\mathbf P}_2= {\mathbf P}$), as well as the scattering angle of the relative
motion dynamics  ${\mathbf Q}={\mathbf R}_1-{\mathbf R}_2$, which is the dynamics we consider. We assume that we study
the real relative motion in a class of coordinate gauges which is regular enough at infinity to ensure the gauge-invariance of $\chi_{\rm real}$.
The requirement \eqref{chi=} then amounts to assuming that the canonical transformation (with generating function $\mathcal G$)
linking the real phase-space (relative) variables ${\mathbf Q}$,  ${\mathbf P}$ 
to the effective ones, say ${\mathbf Q'}$,  ${\mathbf P'}$, is such that, asymptotically ${\mathbf P'} \propto {\mathbf P}$. It is easily
seen that this is actually a consequence of the general form of the generating function
\be
{\mathcal G}({\mathbf Q}, {\mathbf P}')  ={\mathbf Q} \cdot {\mathbf P'} \left[1+  c_{11} \left(\frac{{\mathbf P'}}{\mu c}\right) ^2 + c_{12} \frac{GM}{c^2 R} + \cdots\right]\,,
\ee
which was assumed (and found consistent) in previous (PN-based) EOB works.
Finally, let us note that the usefulness of considering the asymptotically defined, gauge-invariant functional link $\chi_{\rm real} ({\mathcal E}_{\rm real}, J) $ 
between scattering angle and the (asymptotic, conserved\footnote{We recall that we focus here on the conservative dynamics of
two-body systems.}) total energy and angular momentum was emphasized in \cite{Damour:2009sm}, within the context of self-force
studies, and was studied in full numerical relativity in \cite{Damour:2014afa}.

\section{General structure of the Post-Minkowskian expansion of the scattering function} \label{sec4}

As a warm up, and for orientation, let us recall the value of the scattering function $\chi_{\rm real} ({\mathcal E}_{\rm real}, J) $ 
in the Newtonian approximation (Rutherford scattering). It is most simply obtained by starting from the polar-coordinate equation of
a conic, namely $R = p/(1+ e \cos \varphi)$ (valid for all three types of conics). In the hyperbolic case (i.e. when $e>1$) the branches at
infinity corresponds to the roots of $1+ e \cos \varphi=0$. This leads to $\Phi_{\rm scatt}= 2 \, {\rm arcos} \left( -1/e \right)$, or,
equivalently,
\be
\chi^{\rm Newton}=2 \, \arctan \left( \frac1{\sqrt{e^2-1}} \right) \,.
\ee
In the Newtonian approximation, the eccentricity $e$ is linked to the non-relativistic energy $E= {\mathcal E} - (m_1+m_2)c^2$
and the angular momentum $J$ via
\be \label{eccentricity}
e^2= 1 + 2 \, \frac{E}{\mu c^2} \left( \frac{c J}{G m_1 m_2} \right)^2  \equiv 1 + 2 \, \hat E j^2 \,,
\ee
where we have introduced the dimensionless versions of $E$ and $J$ used in most PN works, namely
\be \label{hEj}
\hat E \equiv \frac{E}{\mu c^2} \, ; \ j \equiv \frac{c J}{G m_1 m_2} = \frac{c J}{G \mu M}\,.
\ee
This leads to the following explicit form of the Newtonian scattering function
\bea \label{chiN1}
\chi^{\rm Newton}(E, J) &=&2 \, \arctan \left( \frac1{\sqrt{2 \, \hat E j^2}} \right) \nonumber\\
&=& 2 \, \arctan \left( \frac{G m_1 m_2}{c J}
\sqrt{\frac{\mu c^2}{2 E }} \right) \,,
\eea
or
\be \label{chiN2}
\tan \left(\frac{1}{2} \chi^{\rm Newton}(E, J) \right)= \frac{G m_1 m_2}{c J} 
\sqrt{\frac{\mu c^2}{2 E }} \,.
\ee
Note that the velocity of light $c$ cancells between $\hat E$  and $j^2$, as expected for a Newtonian-level result.

While the PN expansion is an expansion in powers of $\frac1{c^2}$, the PM expansion is an expansion
in powers of $G$. We see on Eqs. \eqref{chiN1} or \eqref{chiN2} that $\chi$ starts (as expected) at order $G$ in a PM expansion.
As $\chi$ is a dimensionless quantity that is a function of the two dimensionless quantities $\hat E$ and $j$
(and of the symmetric mass ratio, $\nu \equiv \mu/M$),
 and as the definitions \eqref{hEj} show that $G$ enters only via $j = cJ/(G m_1 m_2) \propto c/G$,
we see that the PM expansion of (half) the scattering function will be  equivalent to an expansion in powers of $1/j \propto G$, say
\be \label{PMexp}
\frac12 \chi^{\rm }(E, J) = \frac1j \chi_{1}(\hat E, \nu) + \frac1{j^2} \chi_{2}(\hat E, \nu) + \frac1{j^3} \chi_{3}(\hat E, \nu) + \cdots
\ee
Here, $\chi_{1}(\hat E, \nu)/j$  is the first post-Minkowskian (1PM) approximation of (half) the scattering function, etc.

Note that each term in the PM expansion of  the scattering function is a function of the energy (defined for $E >0$). Our main
tool here will be to compute and exploit the {\it exact form} of the 1PM function  $\chi_{1}(\hat E, \nu)$.
As the latter function is, again, a dimensionless function of  the dimensionless quantity $\hat E \propto \frac1{c^2}$, we see that
the re-expansion of $\chi_{1}(\hat E, \nu)$ in powers of $\hat E$ will correspond to the part of the PN expansion
of the scattering function that is proportional to $1/J$. More precisely, remembering that  $\frac1j \propto \frac{G}{c}$, and parametrizing
the non-relativistic energy $E >0$ by a corresponding squared velocity $v_E$, defined via
\be
E \equiv \frac12 \mu v_E^2 \ ; \ v_E \equiv \sqrt{\frac{2 E}{\mu}}\,,
\ee
so that $\hat E = \frac {v_E^2}{2 c^2}$, we see that the structure of the PN expansion of (half) the scattering function must be of the form
\bea \label{chiPN0}
&& \frac{1}{2} \chi^{}(E, J)   \sim \frac{Gm_1 m_2}{c J} \left(\frac{c}{v_E} + \frac{v_E}{c} + \left(\frac{v_E}{c}\right)^3 + \cdots \right) \nonumber\\
&+ &\left( \frac{G m_1 m_2}{c J} \right)^2  \left(0 \left(\frac{c}{v_E} \right)^2 +1 + \left(\frac{v_E}{c}\right)^2 + \cdots \right) 
\nonumber \\
&+& \left( \frac{G m_1 m_2}{c J} \right)^3  \left(\left(\frac{c}{v_E} \right)^3 + \frac{c}{v_E} +\frac{v_E}{c} + \cdots \right) \nonumber \\
&+& \cdots
\eea
where the first line sketches the form of the PN expansion of the 1PM term $\chi_1(\hat E, \nu)$, say (now with coefficients)
\be
\chi_1(\hat E, \nu) = 1 \frac{c}{v_E} + \chi_{11}(\nu)\frac{v_E}{c} + \chi_{13}(\nu)\left(\frac{v_E}{c}\right)^3 + \cdots
\ee
the second line (where, as indicated, the term $\propto (c/v_E)^2$ is missing) 
sketches the form of the PN expansion of the 2PM term $\chi_2(\hat E, \nu)$, say
\be
\chi_2(\hat E, \nu) = \chi_{20}(\nu) + \chi_{22}(\nu) \left(\frac{v_E}{c}\right)^2 + \chi_{24}(\nu) \left(\frac{v_E}{c}\right)^4 + \cdots
\ee
while the third line sketches the form of the PN expansion of the 3PM term $\chi_3(\hat E, \nu)$, say
\be
\chi_3(\hat E, \nu) = \chi_{3,-3}(\nu) \left(\frac{c}{v_E} \right)^3 + \chi_{3,-1}(\nu) \frac{c}{v_E} + \chi_{31}(\nu)\frac{v_E}{c} + \cdots
\ee

Note the presence of {\it inverse powers} of the velocity $v_E$, especially in the terms odd in $J$ (and in $G$). [Such terms
will also appear in the even contributions $\chi_{2n}(\hat E, \nu)$, where they start at order $(c/v_E)^{2n -2}$.] 
Note also that the coefficients of all the contributions $\propto (G m_1 m_2/cJ)^n (c/v_E)^n$ will vanish for $n>1$ if one
considers the function $\tan (\chi/2)$ instead of $\chi/2$.

Each term in the usual  PN expansion of $\chi/2$  is obtained from the expansion \eqref{chiPN0} above by collecting all 
the contributions $\sim (G m_1 m_2/cJ)^n (v_E/c)^k$ carrying a given power of $\frac1c$, i.e. having a given value of the PN order $\frac12 (n+k)$. The presence of negative powers of $v_E$ (up to $\sim v_E^{-n} $ or $\sim v_E^{-n+ 2}$ in $\chi_{n}$) implies that each term of the PN
expansion (generally) collects an infinite number of contributions of the PM expansion. [This infinite number of contributions corresponds
to the eccentricity dependence of a given PN contribution to $\chi/2$. Indeed, in view of Eq. \eqref{eccentricity}, the eccentricity 
is of order zero in $\frac1{c^2}$, but of order $j^1$ when considering the large $j$ expansion \eqref{PMexp} which defines the 
PM expansion of the scattering function.] For instance the 1PN [i.e. $O(\frac1{c^2}$)] contribution to the scattering function would read
\bea
\frac{1}{2} \chi^{}(E, J)_{1PN}&=  \chi_{11}(\nu) \frac{Gm_1 m_2}{c J} \frac{v_E}{c} + \chi_{20}(\nu) \left( \frac{G m_1 m_2}{c J} \right)^2 \nonumber \\
&+ \chi_{3,-1}(\nu) \left( \frac{G m_1 m_2}{c J} \right)^3 \frac{c}{v_E} + \cdots
\eea
The coefficients $\chi_{nk}(\nu)$ are polynomials in $\nu$ whose
degrees linearly increase with (and seems to be generally equal to) the PN order $\frac12 (n+k)$ of $(G m_1 m_2/cJ)^n (v_E/c)^k$. [See below for examples of these polynomials.]

Let us mention in this respect that the PN expansion of the scattering function has been computed to 2PN accuracy in section 5C
of Ref. \cite{Bini:2012ji}. From the results there one can then deduce (by collecting the coefficient of the
first power of $1/j$ in $\frac{1}{2} \chi^{}(E, J)_{1PN}$) the beginning of the $v_E$ expansion of the 1PM term $\chi_1(\hat E, \nu)$.
One then finds
\bea \label{chi12PN}
\chi_1({\hat E}_{\rm real}, \nu)&=  \frac{c}{v_E^{\rm real}} + \frac{15 -\nu}{8}\frac{v_E^{\rm real}}{c} + \frac{35 + 30 \nu + 3 \nu^2}{128}\left(\frac{v_E^{\rm real}}{c}\right)^3 \nonumber\\
&+ O( (\frac{v_E^{\rm real}}{c} )^5) \,.
\eea
Here we have specified that the energy used in this result to express the PN-expanded 1PM contribution to the scattering angle
is the dimensionless $\mu c^2$-rescaled {\it real} non-relativistic energy 
\be \label{defEreal}
{\hat E}_{\rm real}= \frac{{E}_{\rm real}}{\mu c^2}=\frac{{\mathcal E}_{\rm real} - Mc^2}{\mu c^2}\,,
\ee
the auxiliary ``velocity" entering \eqref{chi12PN} denoting simply 
\be \label{defvE}
v_E^{\rm real} \equiv c \sqrt{2 {\hat E}_{\rm real}} \equiv \sqrt{\frac{2  E_{\rm real}}{\mu}}.
\ee
[In the above discussion, we had left unspecified whether we were dealing with $\chi_{\rm real}$ as a function of $E_{\rm real}$
or with $\chi_{\rm eff}$ as a function of $E_{\rm eff}$.]

The result \eqref{chi12PN} illustrates in advance the {\it complementarity} between the PM-expansion approach used in the
present paper and the PN expansions used in previous works. Each term of the PM
expansion collects an infinite number of contributions of the PN expansion\footnote{This is the reciprocal of the fact mentioned
above that each term of the PN expansion collects an infinite number of contributions of the PM expansion.}.
Therefore, the exact computation of a certain term in the PM expansion of $\chi$ (as we shall report below) represents
an information about the two-body dynamics which goes beyond all the present PN knowledge (which is limited to
the 4PN level \cite{Damour:2014jta}, \cite{Damour:2015isa}), though it evidently misses some of the information contained
in the PN computations of $\chi$.

\section{Real, two-body scattering function at the first post-Minkowskian approximation} \label{sec5}

The relativistic gravitational two-body scattering function $\frac12 \chi_{\rm real}( {\mathcal E}_{\rm real}, J; m_1, m_2)$ (now
expressed in terms of the total relativistic energy, $ {\mathcal E}_{\rm real}=   (m_1+m_2) c^2 + E_{\rm real}$) has been computed at the first post-Minkowskian approximation
(i.e. first order in $G$) by several authors \cite{Portilla:1979xx,Portilla:1980uz,Westpfahl:1985,Westpfahl:1987,Ledvinka:2008tk}. We recall that ${\mathcal E}_{\rm real}$, and $J$, are both evaluated in the {\it cm frame} of the two-body system.
In the following, it will often be convenient to use units where $c=1$.

The results for $\chi_{ 1PM}$ given in the articles cited above are not written in a way which highlights the basic physics underlying the
two-body scattering. We shall therefore give a novel, more transparent derivation of  $\chi_{ 1PM}$, which also exhibits the link between the
classical scattering function $ \chi_{\rm real}( {\mathcal E}_{\rm real}, J; m_1, m_2)$ and the quantum scattering  two-body
amplitude $\langle p'_1 p'_2 | p_1 p_2 \rangle$. Both results will appear to be directly deducible from the diagram displayed in Fig. 1.

\begin{figure}[t]
\includegraphics[height=5cm]{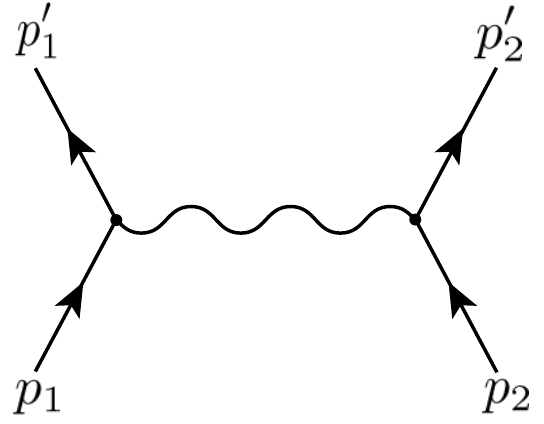}
\caption{\label{fig:1}Diagram displaying the physical ingredients of both the classical and the quantum  two-body scattering.}
\end{figure}

The gravitational equation of motion of, say, particle 1, is most simply written as
\be \label{eom1}
\frac{d p_{1 \mu}}{d \sigma_1} = \frac12 \,\D_\mu g_{\alpha \beta}(x_1) \, p_1^{\alpha} p_1^{\beta}\,.
\ee
Here, $p_{1 \mu}= g_{\mu \nu}(x_1)\, p_1^{\nu}$ denotes the (curved spacetime) covariant components of $p_1^{\mu} = m_1 d x_1^{\mu}/ds_1$, where $ds_1$ denotes the proper time along the worldline $x_1^{\mu}=x_1^{\mu}(s_1)$ of $m_1$. [We use a mostly positive signature with, e.g., $ g^{\mu \nu} p_{1 \mu} p_{1 \nu} = - m_1^2$.]
We have also introduced the rescaled
proper time $\sigma_1$ defined as $\sigma_1= s_1/m_1$ so that $p_1^{\mu} = d x_1^{\mu}/d\sigma_1$. Integrating \eqref{eom1}
with respect to  $\sigma_1$ (between $-\infty$ and $+\infty$) yields the total change 
$ \Delta p_{1 \mu} \equiv p'_{1 \mu} - p_{1 \mu}$ in the asymptotic 4-momentum of particle 1:
\be \label{dp1}
\Delta p_{1 \mu} = \int_{-\infty}^{+\infty} d \sigma_1 \frac12 \, p_1^{\alpha} p_1^{\beta} \,\D_\mu h_{\alpha \beta}(x_1) \, ,
\ee
where $h_{\alpha \beta} \equiv g_{\alpha \beta} - \eta_{\alpha \beta}$.

At linearized order in $G$,  $p_1^{\alpha}$ on the right-hand side (rhs) of \eqref{dp1} can be replaced by the constant incoming 4-momentum of particle 1,
while the metric perturbation can be replaced by the solution of the linearized Einstein equations, namely (in four spacetime dimensions;
and in harmonic gauge)
\be
\Box h_{\alpha \beta} = - 16 \pi G \left(T_{\alpha \beta} - \frac12 T \eta_{\alpha \beta} \right) \,.
\ee
Here and below, all index operations are performed in the flat Minkowski background (e.g. $T = \eta_{\alpha \beta} T^{\alpha \beta}$).
At order $G$, only the metric perturbation generated by the (flat spacetime) stress-energy tensor of particle 2,
\be
T_2^{\alpha \beta}(x)=  \int_{-\infty}^{+\infty} d \sigma_2  \, p_2^{\alpha} p_2^{\beta} \, \delta^4(x-x_2(\sigma_2))\,,
\ee
needs to be inserted in the computation \eqref{dp1} of $\Delta p_{1 \mu} $.

In order to exhibit the link between the classical scattering and the usual Feynman diagram corresponding to Fig. 1, let us work
in (4-dimensional) Fourier space ($k.x \equiv k_{\mu} x^{\mu} \equiv \eta_{\mu \nu} k^{\mu} x^{\nu}$),
\be
h_{\alpha \beta}(x)= \int \frac{d^4 k}{(2\pi)^4} h_{\alpha \beta}(k) e^{i k.x}\,.
\ee
The $m_2$-generated metric perturbation reads
\be \label{h2x}
h_{2 \alpha \beta}(x)= 16 \pi G \int \frac{d^4 k}{(2\pi)^4} e^{i k.x} \frac{P_{\alpha \beta ;\alpha' \beta'}}{k^2} T_2^{ \alpha' \beta'}(k) \,,
\ee
where $P_{\alpha \beta ;\alpha' \beta'}/k^2$, with $P_{\alpha \beta ;\alpha' \beta'}\equiv \eta_{\alpha \alpha'} \eta_{\beta \beta'} -\frac12 \eta_{\alpha \beta} \eta_{\alpha' \beta'}$, and $k^2 \equiv \eta^{\mu \nu} k_{\mu} k_{\nu}$, is the Fourier-space gravitational propagator. [At this order, it does not matter whether one considers a retarded or a time-symmetric propagator (with $1/k^2$ then denoting a principal value kernel).]

The $k$-space stress-energy tensor of $m_2$ reads
\bea \label{T2k}
T_2^{ \alpha' \beta'}(k)&=& \int d^4 x e^{- i k.x} T_2^{ \alpha' \beta'}(x) \nonumber \\
&=&  \int_{-\infty}^{+\infty} d \sigma_2   e^{- i k.x_2(\sigma_2)}  \, p_2^{\alpha'} p_2^{\beta'}\,.
\eea
Inserting, successively, \eqref{T2k} into \eqref{h2x}, and \eqref{h2x} into \eqref{dp1}, leads to the following explicit expression
for the total change of 4-momentum of particle 1
\bea \label{dp1'}
\Delta p_{1 \mu}= 8 \pi G \int  \frac{d^4 k}{(2\pi)^4} i k_{\mu} p_1^{\alpha} p_1^{\beta} \frac{P_{\alpha \beta ;\alpha' \beta'}}{k^2} p_2^{\alpha'} p_2^{\beta'} \nonumber \\ 
\times \int d \sigma_1 \int d \sigma_2   e^{ i k.(x_1(\sigma_1) - x_2(\sigma_2))}\,.
\eea
[By changing $p_1 \to p_2, x_1 \to x_2, k\to -k$, one immediately sees that one would have $\Delta p_{2 \mu} = -\Delta p_{1 \mu}$.]
On the first line we recognize all the ingredients of the quantum scattering amplitude of Fig. 1: the two matter-gravity vertices $p_1^{\alpha} p_1^{\beta}$
and $p_2^{\alpha'} p_2^{\beta'} $ (computed in the approximation $p'_1\approx p_1$, $p'_2\approx p_2$), connected by the gravitational propagator $P_{\alpha \beta ;\alpha' \beta'}/k^2$. At first order in $G$, all the ingredients entering the rhs of \eqref{dp1'} can be replaced
by their zeroth-order (free motion) approximations, i.e. constant momenta (as already used) and straight (incoming) worldlines:
\be \label{worldlines}
x_1^{\mu}(\sigma_1)= x_1^{\mu}(0) + p_1^{\mu} \sigma_1 \, ; \, x_2^{\mu}(\sigma_2)= x_2^{\mu}(0) + p_2^{\mu} \sigma_2 \,.
\ee

Inserting the straight-worldlines expressions \eqref{worldlines} into Eq. \eqref{dp1'} allows one to explicitly compute the $\sigma_1$
and $\sigma_2$ integrals on the second line, with the result
\be \label{secondline}
(2 \pi)^2  e^{ i k.\left(x_1(0) - x_2(0)\right)} \delta( k.p_1)  \, \delta(k.p_2) \,.
\ee
The crucial point is that the $\sigma_1$ and $\sigma_2$ integrals have generated two (one-dimensional) delta functions involving
two different linear combinations of the four Fourier-space variables $k_{\mu}$. This restricts the 4-dimensional integral over $k_{\mu}$
appearing on the first line of  Eq. \eqref{dp1'},
\be
\int  \frac{d^4 k}{(2\pi)^4} \frac{i k_{\mu}}{k^2} \, (\cdots)\,,
\ee
 to a two-dimensional linear subspace of $k$-space.
 
We can explicitly deal with this two-dimensional reduction of the $k$-integral by choosing an adapted (Lorentz) coordinate
frame. More precisely, it is convenient to choose four (Lorentz-orthonormal) 4-vectors $ e_0, e_1, e_2, e_3$ such that,
say, $e_0$ and $e_3$ span the (timelike) two-plane defined by the two 4-vectors $p_1$ and $p_2$.  This will be, in particular,
the case, if we work in a cm frame with time axis defined by $e_0 \propto p_1+ p_2$, and third axis $e_3$ in the common
direction of  the cm  spatial momenta, say ${\bf p}_{cm} \equiv {\bf p_1}^{cm} = - {\bf p_2}^{cm}$ . In this frame, the incoming 4-momenta have
the following components:
\be
p_1= {\mathcal E}_1^{cm} e_0 +  p_{cm} e_3 \, ; \, p_2= {\mathcal E}_2^{cm} e_0 -  p_{cm} e_3\,,
\ee
where $p_{cm}$ is the magnitude of ${\bf p}_{cm}$, and where  ${\mathcal E}_1^{cm} = \sqrt{m_1^2 + p_{cm}^2}$
and ${\mathcal E}_2^{cm} = \sqrt{m_2^2 + p_{cm}^2}$ are the relativistic  cm energies of the two (incoming) particles.
In terms of these quantities, the total relativistic (incoming, center-of-mass) energy reads
\be \label{Ecm}
{\mathcal E}_{\rm real}= {\mathcal E}_1^{cm} + {\mathcal E}_2^{cm}= \sqrt{m_1^2 + p_{cm}^2} + \sqrt{m_2^2 + p_{cm}^2}\,.
\ee
Among the four components of $k$ in the frame $ e_0, e_1, e_2, e_3$, only $k^0$ and $k^3$ appear in the two delta functions
of Eq. \eqref{secondline}. More precisely they appear in the combinations
\be\label{kp1kp2}
k.p_1= - k^0  {\mathcal E}_1^{cm} + k^3 p_{cm}\, ;\, k.p_2= - k^0  {\mathcal E}_2^{cm} - k^3 p_{cm}\,.
\ee
As a consequence, we can write
\be
\delta( k.p_1)  \, \delta(k.p_2)  =  \frac{\delta( k^0)  \delta(k^3) }{\mathcal D}\,,
\ee
where ${\mathcal D}$ is the absolute value of the Jacobian $\D(k.p_1,k.p_2)/\D(k^0,k^3)$.
We can immediately give two different expressions for the Jacobian ${\mathcal D}$. First, from Eq. \eqref{kp1kp2} we get
\be \label{D1}
{\mathcal D} = \left(  {\mathcal E}_1^{cm} +  {\mathcal E}_2^{cm} \right) p_{cm} = {\mathcal E}_{\rm real} \, p_{cm} \,.
\ee
Second, we can write a covariant expression for ${\mathcal D}$ by thinking geometrically within the two-dimensional Lorentzian
space spanned by $p_1$ and $p_2$, and realizing that ${\mathcal D}$ is simply the magnitude of the wedge product (antisymmetric
bi-vector) $p_1 \wedge p_2$ of $p_1$ and $p_2$. This yields the manifestly covariant expression (with a minus sign linked
to the timelike character of the $p_1$-$p_2$ plane)
\bea \label{D2}
{\mathcal D}^2 &=& |p_1 \wedge p_2|^2 = -\frac12 (p_1^{\mu} p_2^{\nu}- p_1^{\nu} p_2^{\mu}) (p_{1\mu} p_{2\nu}- p_{1\nu} p_{2\mu}) \nonumber \\
&=& (p_1.p_2)^2 - p_1^2 \, p_2^2 \,.
\eea
Note in passing that the (so proven) equality between ${\mathcal E}_{\rm real} \, p_{cm} $ and $\sqrt{(p_1.p_2)^2 - p_1^2 \, p_2^2}$
is not evident when using the explicit cm expression \eqref{Ecm} of the cm energy ${\mathcal E}_{\rm real}$ in terms of $p_{cm}$.

Inserting our various partial results in the integral expression \eqref{dp1'} we get a result of the form
\be \label{dp1''}
\Delta p_{1 a} \propto G \, \frac{p_1^{\alpha} p_1^{\beta} P_{\alpha \beta ;\alpha' \beta'} p_2^{\alpha'} p_2^{\beta'} }{{\mathcal D}} \int d^2k \frac{ i k_a}{{\bf k}^2} e^{i {\bf k} . {\bf b}} \,,
\ee
where $\bf k$ and $\bf b$ denote the projections, onto the two-dimensional (Euclidean-signature) space spanned by $e_1$ and $e_2$,
of $k$ and $x_1(0) - x_2(0)$. [The index $a = 1, 2$ on $\Delta p_{1 a}$ and $k_a$ spans this two-dimensional space; e.g. $k^a e_a = k^1 e_1+ k^2 e_2$.] Clearly, $\bf b$ represents the vectorial cm impact parameter of the two incoming worldlines, and its Euclidean
magnitude $b \equiv |\bf b|$ measures the scalar cm impact parameter. As a consequence, the cm total angular momentum is simply
given by
\be
J = b \, p_{cm} \,.
\ee
From dimensional analysis and symmetry considerations (or by an explicit calculation, using a frame where, say, $b^2=0$), the remaining two-dimensional integral in \eqref{dp1''}
is simply found to be proportional to $ - {\bf b} /b^2$. [The vector ${\bf b} = {\bf x}_1(0) - {\bf x}_2(0)$ is directed from 2 towards 1.]

Putting together all the numerical factors, one finally gets a vectorial deflection (within the $e_1$-$e_2$ two-plane, i.e. the 
orthogonal complement of the  $p_1$-$p_2$ two-plane) given by
\be
\Delta {\bf p}_1= - 4 G \frac{p_1^{\alpha} p_1^{\beta} P_{\alpha \beta ;\alpha' \beta'} p_2^{\alpha'} p_2^{\beta'} }{{\mathcal D}} \frac{\bf b}{b^2} \,.
\ee
The (absolute value of the) corresponding cm scattering angle $\chi$ is related to the magnitude of $\Delta {\bf p}_1 = - \Delta {\bf p}_2$
via 
\be
\sin \frac{\chi}{2}=\frac{ |\Delta {\bf p}_1|}{2 \,  |{\bf p}_1|} = \frac{ |\Delta {\bf p}_1|}{2 \, {p}_{cm}}\,.
\ee
As we work to first order in $G$ we have $\sin \frac{\chi}{2} \approx \frac{\chi}{2}$ so that
\bea\label{chireal1}
\frac12 \chi^{\rm real}_{1PM} &=& 2 \frac{ G }{ b\,  p_{cm} } \frac{ p_1^{\alpha} p_1^{\beta} P_{\alpha \beta ;\alpha' \beta'} p_2^{\alpha'} p_2^{\beta'} }{{\mathcal D}} \nonumber \\
&=& 2 \frac{ G }{ J } \frac{ p_1^{\alpha} p_1^{\beta} P_{\alpha \beta ;\alpha' \beta'} p_2^{\alpha'} p_2^{\beta'} }{{\mathcal D}} \,.
\eea
In the non-relativistic limit, $p_1^{\alpha} p_1^{\beta} P_{\alpha \beta ;\alpha' \beta'} p_2^{\alpha'} p_2^{\beta'} = (p_1.p_2)^2 - \frac12 p_1^2 p_2^2$ tends to $\frac 12 m_1^2 m_2^2$, where the factor $\frac 12$ cancells the prefactor $2$ on the rhs of \eqref{chireal1}.
[In agreement with the Newtonian-level result \eqref{chiN1}.]

The expression \eqref{chireal1} (and its derivation above) exhibits a simple, and manifestly covariant, link between the (Fourier-space) quantum scattering amplitude
${\mathcal M} = G p_1^{\alpha} p_1^{\beta} P_{\alpha \beta ;\alpha' \beta'} p_2^{\alpha'} p_2^{\beta'}/k^2$ and the classical
scattering angle.

The explicit form of Eq. \eqref{chireal1} reads
\be \label{chireal2}
\frac12 \chi^{\rm real}_{1PM} = \frac{ G }{ J }  \frac{ 2 (p_1.p_2)^2 - p_1^2 p_2^2 }{ \sqrt{(p_1.p_2)^2 - p_1^2 p_2^2} }\,.
\ee
This result naturally features the following dimensionless energy variable (already used in Ref. \cite{Damour:1997ub})
\be \label{epsilon}
\epsilon(s) \equiv \frac{s - m_1^2 -m_2^2}{2 m_1 m_2} = - \frac{p_1.p_2}{ m_1 m_2}\,.
\ee
Here $s$ denotes the usual Mandelstam variable
\be
s \equiv {\mathcal E}_{\rm real}^2 = - (p_1+p_2)^2\,,
\ee
where we recall that $ {\mathcal E}_{\rm real}$ is evaluated in the cm.

In terms of $\epsilon(s)$, the result \eqref{chireal2} reads
\be \label{chireal3}
\frac12 \chi^{\rm real}_{1PM}(s, J)=  \frac{ G m_1 m_2}{ J }  \frac{ 2 \, \epsilon^2(s) - 1 }{ \sqrt{\epsilon^2(s) - 1} }\,.
\ee
Our explicit final result \eqref{chireal2} [or \eqref{chireal3}] is simpler than the corresponding results derived (in $x$-space) in
Refs. \cite{Portilla:1979xx,Portilla:1980uz,Westpfahl:1985,Westpfahl:1987,Ledvinka:2008tk}. It is, however, 
equivalent to them. [We note in passing that the equivalence with Eq. (12) in \cite{Ledvinka:2008tk} is rather hidden.]
The expression \eqref{chireal3} is also simpler than the corresponding 2PN-expanded result \eqref{chi12PN}, but is straightforwardly
checked to be consistent with it, when remembering the definitions \eqref{defEreal}, \eqref{defvE}.

\section{Effective one-body scattering function at the first post-Minkowskian approximation} \label{sec6}

In this section we shall consider the (effective) scattering angle $\chi_{\rm eff}$ computed from the dynamics of one
particle of mass $m_0$ moving in some  effective metric $g_{\mu \nu}^{\rm eff}$. At linear order in $G$
(and still setting $c=1$), we parametrize the looked-for spherically symmetric effective metric as 
\bea \label{geffbeta1}
g_{\mu \nu}^{\rm eff}(M_0, \beta_1)dx^{\mu} dx^{\nu} &=& - \left(1- \frac{R_g}{R}\right) \,dt^2 + \left(1+ \beta_1 \frac{R_g}{R} \right) \,dR^2 \nonumber\\
&+& R^2 \left(d\theta^2 + \sin^2\theta \, d \varphi^2\right)\,,
\eea
where the dimensionless parameter $\beta_1$ enters in the radial component of the metric, and where
\be
R_g \equiv 2 G M_0\,,
\ee
denotes the (gravitational) mass of the effective background. We shall set below  $m_0$ to the reduced mass $\mu$ and
$M_0$ to the total mass $M = m_1+m_2$, but it is useful not to assume it from the start.

Let us compute the scattering angle $\chi_{\rm eff}$ from the Hamilton-Jacobi equation for the geodesic dynamics in $g_{\mu \nu}^{\rm eff}(M_0)$, namely
\be
g^{\mu \nu}_{\rm eff} \, \D_{\mu} S_{\rm eff}  \, \D_{\mu} S_{\rm eff} = - m_0^2 \,,
\ee
with
\be
S_{\rm eff}= - {\mathcal E}_0 t + J_0 \varphi + S^{\rm eff}_R(R)\,.
\ee
For simplicity, we denote the effective energy and angular momentum  ${\mathcal E}_{\rm eff} $ and $ J_{\rm eff}$ 
as ${\mathcal E}_0 $ and $ J_0$, respectively.

To linear order in $G$, the Hamilton-Jacobi equation reads
\be
- \left(1 + \frac{R_g}{R}\right) {\mathcal E}_0^2 + \frac{J_0^2}{R^2} + \left(1- \beta_1 \frac{R_g}{R}\right) \left(\frac{d S^{\rm eff}_R}{dR}\right)^2= -m_0^2 \,.
\ee
Computing $P_R= d S^{\rm eff}_R/dR$ from this equation, we obtain (still to linear order in $G$) the radial effective action in the form
\be
S^{\rm eff}_R(R; {\mathcal E}_0, J_0)= \int dR P_R(R; {\mathcal E}_0, J_0) \,,
\ee
with
\bea
 P_R(R; {\mathcal E}_0, J_0) &=& \pm \left(1+ \frac12 \beta_1 \frac{R_g}{R}\right) \nonumber\\
 &\times& \sqrt{{\mathcal E}_0^2 \left(1 + \frac{R_g}{R}\right)- \left( m_0^2 +  \frac{J_0^2}{R^2} \right)    } \,,
\eea
where the sign $\pm$ is $-$ (respectively, $+$) during the incoming (resp. outgoing) part of the scattering motion.

As already explained in Section \ref{sec3}, the $J_0$ derivative of the radial action directly yields the scattering angle as
\be \label{chieff0}
\pi + \chi_{\rm eff}= -\int_{- \infty}^{+\infty} dR \frac{\partial P_R(R; {\mathcal E}_0, J_0)}{\partial J_0} \, .
\ee
Let us formally expand  $P_R$ in powers of $G$, i.e. of $R_g$:
\bea
P_R(R; {\mathcal E}_0, J_0; R_g) &=& P_R^{(0)}(R; {\mathcal E}_0, J_0) +  P_R^{(1)}(R; {\mathcal E}_0, J_0) \nonumber \\
&+& O(R_g^2) \,.
\eea
Here, $ P_R^{(0)}=  \sqrt{{\mathcal E}_0^2 - \left( m_0^2 +  \frac{J_0^2}{R^2} \right)}$ corresponds to a free motion and
therefore contributes the no-scattering value $\pi$ to the integral \eqref{chieff0}. We then get the following expression for
the (effective) scattering angle
\be
\chi_{\rm eff}= -\int_{- \infty}^{+\infty} dR \frac{\partial P_R^{(1)}(R; {\mathcal E}_0, J_0)}{\partial J_0}\,.
\ee
Actually, this expression is formal because, when explicitly written, it involves divergent integrals linked to the singular
nature of the expansion of $\sqrt{{\mathcal E}_0^2 - \left( m_0^2 +  \frac{J_0^2}{R^2} \right) + \varepsilon}$ in powers
of $\varepsilon$ at the turning point where $P_R$, i.e. the squareroot, vanishes. However, as explained in Ref. \cite{Damour:1988mr},
the correct result is obtained by taking the Hadamard partie finie (Pf), at the unperturbed turning point, of these formally
divergent integrals. This yields
\be
\chi_{\rm eff}=\frac{R_g J_0}{2} \, {\rm Pf} \left[\int \pm \frac{ dR}{R^3} \left( \beta_1 A^{-1/2} - {\mathcal E}_0^2 A^{-3/2} \right) \right]\,,
\ee
where $A \equiv  {\mathcal E}_0^2 - \left( m_0^2 +  J_0^2/R^2\right)$.

The integrals entering this expression of $\chi_{\rm eff}$ are elementary to compute when replacing $R$ by $A$ as integration
variable (using ${\rm Pf} \int_{\rm in}^{\rm out} \pm dA A^p =  \frac1{p+1} (A_{\rm out}^{p+1}+ A_{\rm in}^{p+1})$ because of the two signs of the squareroot of $A$). A simple computation then yields
\bea \label{chieff2}
\frac 12 \chi^{\rm eff}_{1PM}({\mathcal E}_0, J_0)&=& \frac{G M_0}{J_0} \left( \beta_1 \sqrt{ {\mathcal E}_0^2 -  m_0^2}  + \frac{{\mathcal E}_0^2}{\sqrt{ {\mathcal E}_0^2 -  m_0^2} } \right) \nonumber\\
&=& \frac{G M_0}{J_0} \, \frac{ (1+\beta_1 ) \, {\mathcal E}_0^2   - \beta_1 \, m_0^2}{\sqrt{ {\mathcal E}_0^2 -  m_0^2} } \,.
\eea
Factoring $m_0$ out of the effective energy ${\mathcal E}_0$ yields the equivalent expression
\be \label{chieff3}
\frac 12 \chi^{\rm eff}_{1PM}({\mathcal E}_0, J_0) =  \frac{G M_0 m_0}{J_0} \, \frac{ (1+\beta_1 ) \, ({\mathcal E}_0/m_0)^2   - \beta_1 }{\sqrt{ ({\mathcal E}_0/m_0)^2 -  1} }\,.
\ee
We recall that, in the expressions above, ${\mathcal E}_0 $ denotes the relativistic effective energy ${\mathcal E}_{\rm eff} $, 
and $ J_0$ the effective angular momentum $ J_{\rm eff}$.

\section{Comparing the real and the effective scattering functions} \label{sec7}

We have recalled above that the EOB formalism is a relativistic generalization of the  Newtonian
fact that the relative dynamics of a two-body system is  (after separation of the center of mass motion) equivalent
to the motion of a particle of mass $\mu$ submitted to the original two-body potential $V(| {\bf R}_1 -  {\bf R}_2|)$.
Therefore, in the non-relativistic limit $c \to \infty$, we require that the effective mass $m_0$ coincides with $\mu$, and that,
as already indicated in Eq. \eqref{fexp}, the non-relativistic effective energy $E_{\rm eff} = {\mathcal E}_{\rm eff} - m_0 c^2$
coincide with the non-relativistic real cm energy $E_{\rm real} = {\mathcal E}_{\rm real} - (m_1+m_2) c^2$. In this non-relativistic
limit, the real and effective scattering functions, evaluated for the same values of their arguments, $E_{\rm real} = E_{\rm eff}$,
and $J_{\rm real} = J_{\rm eff}$ [see Eq. \eqref{Jdic}], must coincide: $\chi_{\rm eff}(E, J) = \chi_{\rm real}(E, J) + O(\frac1{c^2})$.
Applying this requirement to Eqs. \eqref{chireal3} and \eqref{chieff3} simply tells us that 
\be \label{M0m01}
M_0 m_0 = m_1 m_2 + O\left(\frac1{c^2}\right) \,.
\ee
Though other possibilities have been explored in \cite{Buonanno:1998gg}, it was concluded that it is most natural to work
with effective masses $m_0$ and $M_0$ which are energy-independent. The requirements  $m_0=\mu + O(\frac1{c^2})$
and \eqref{M0m01} then imply that
\be \label{M0m02}
m_0 = \mu \equiv \frac{m_1 m_2}{m_1+m_2} \ ; \  M_0= M \equiv m_1 + m_2 \,.
\ee
Then, the application of our basic scattering-state dictionary \eqref{dic1} [or \eqref{dic2}] to our scattering results \eqref{chireal3},
\eqref{chieff3} imply that the a priori unknown energy map $f$, Eq. \eqref{energymap} [such that
${\mathcal E}_0 = f({\mathcal E}_{\rm real}) \equiv f(s)$], as well as the a priori unknown effective metric parameter
$\beta_1$ [see Eq. \eqref{geffbeta1}]  should be such that
\be \label{requirement1}
\frac{ (1+\beta_1 ) \, (f(s)/m_0)^2   - \beta_1 }{\sqrt{ (f(s)/m_0)^2 -  1} }=   \frac{ 2 \, \epsilon^2(s) - 1 }{ \sqrt{\epsilon^2(s) - 1} } \,,
\ee
where the function $\epsilon(s)$ of the real cm energy was defined in Eq. \eqref{epsilon}.

The requirement \eqref{requirement1} contains both an unknown parameter, $\beta_1$, and an unknown function, $f(s)$. It would
a priori seem that this is not enough to determine all the unknowns. One could think that one can choose an arbitrary value
of $\beta_1$ and then determine the corresponding energy map $f(s)$ by solving Eq.  \eqref{requirement1} for $f(s)$.
Let us show, however, that the exact, relativistic structure of \eqref{requirement1} is such that it uniquely determines {\it both}
the value of $\beta_1$, and that of the energy map $f(s)$.

Indeed, denoting $u_{\rm eff} \equiv \sqrt{ (f(s)/m_0)^2 -  1}$ and, correspondingly, $u_{\rm real} \equiv \sqrt{\epsilon^2(s)- 1}$,
the left-hand side of Eq. \eqref{requirement1} has the structure
\be
\frac{ (1+\beta_1 ) \, (u_{\rm eff}^2 +1)   - \beta_1 }{u_{\rm eff}} = (1+\beta_1 ) \, u_{\rm eff} + \frac1{u_{\rm eff}}   \,,
\ee
while its rhs  has the structure
\be
\frac{ 2 \, (u_{\rm real}^2 +1)   - 1 }{u_{\rm real}} = 2 \, u_{\rm real} + \frac1{u_{\rm real}}\,.
\ee
However, the variable $u_{\rm real}$ runs over the full half-line $0 \leq u_{\rm real} \leq + \infty$ (with the limit $u_{\rm real} \to 0$
corresponding to the non-relativistic limit), while $u_{\rm eff} \geq 0$  must also start at $0$ in the non-relativistic limit.
Now, a remarkable feature of the function $g_{\rm real}(u_{\rm real}) =2 \, u_{\rm real} + \frac1{u_{\rm real}}$ (which describes the energy dependence
of the product $\chi J$) is that, after {\it initially decreasing} with increasing energy in the non-relativistic regime ($u_{\rm real}\approx v_E \ll 1$
implying $g_{\rm real} \approx 1/v_E = 1/\sqrt{2 \hat E}$; Rutherford scattering), it ultimately starts to {\it increase} with increasing energy
in the relativistic energy domain ($g_{\rm real} \sim  u_{\rm real} \propto s= {\mathcal E}_{\rm real}^2$). Therefore, the function $g_{\rm real}(u_{\rm real})$ is easily found to have a (unique) minimum. The minimum value of $g_{\rm real}(u_{\rm real}) =2 \, u_{\rm real} + \frac1{u_{\rm real}}$ is reached for $u_{\rm real}= \frac1{\sqrt{2}}$ and is equal to $ 2 \sqrt{2}$.

By comparison a general function of the type $(1+\beta_1 ) \, u_{\rm eff} + \frac1{u_{\rm eff}} $,
where the variable $u_{\rm eff}$ lives on the positive half-real-line, must have $1+ \beta_1 \geq 0$ to remain
always positive, and will then reach the minimal value $ 2 \sqrt{1+\beta_1}$ for $u_{\rm eff} = 1/\sqrt{1+\beta_1}$.

In order for the requirement  \eqref{requirement1} to be globally satisfied, in the relativistic regime, we must identify
the two minimum values $ 2 \sqrt{2}$ and $ 2 \sqrt{1+\beta_1}$. We therefore conclude that the value of $\beta_1$
is uniquely determined to be
\be \label{beta1}
\beta_1= 1\,,
\ee
which corresponds to the linearized Schwarzschild metric.

Inserting the information \eqref{beta1} in the requirement  \eqref{requirement1} one can now conclude that there exists a {\it unique}
energy map which is positive, continuous and monotonic, and that it is given by
\be
\frac{f(s)}{m_0} = \epsilon(s)\equiv \frac{s - m_1^2 -m_2^2}{2 m_1 m_2} \,,
\ee
i.e.,  using \eqref{M0m02},
\be \label{fx}
\frac{{\mathcal E}_{\rm eff}}{\mu} =  \frac{({\mathcal E}_{\rm real})^2 - m_1^2  -m_2^2 }{2 m_1 m_2} \,,
\ee
or, equivalently,
\be \label{fEOB1}
{\mathcal E}_{\rm eff}= \frac{({\mathcal E}_{\rm real})^2 - m_1^2  -m_2^2 }{2 \, (m_1+m_2) } \, .
\ee
The PN expansion, \eqref{fexp}, of this map yields $\alpha_1=\frac{\nu}{2}$ and $\alpha_n\equiv 0$ for all higher $n$'s.

\section{$O(G)$ tensor-scalar generalization}  \label{sec8}

To further illustrate the usefulness of the post-Minkowskian scattering approach, let us briefly consider the modifications
of the EOB results brought by considering a gravitational interaction combining massless spin-2 exchange and massless spin-0
exchange. In diagrammatic language, this amounts to considering the sum of two one-particle-exchange diagrams of the form of Fig. 1.
At first order in the interaction (and therefore neglecting self-field effects), the scattering angle is then simply the sum of two contributions
\be
\chi=\chi_2 + \chi_0 \,.
\ee
The spin-2 contribution $\chi_2$ can be written in terms of the  4-velocities $u_1 \equiv p_1/m_1$ and $u_2 \equiv p_2/m_2$
[with $\widehat {\mathcal D} \equiv {\mathcal D}/(m_1m_2) = \sqrt{(u_1.u_2)^2 -1}$], as
\be \label{chispin2}
\frac12 \chi_2=   \frac{ G_2 m_1 m_2 }{ J }\,  \frac{ 2 \, u_1^{\alpha} u_1^{\beta} P_{\alpha \beta ;\alpha' \beta'} u_2^{\alpha'} u_2^{\beta'} }{\widehat {\mathcal D}} \,.
\ee
 The spin-0 contribution $\chi_0$ is then simply  obtained by omitting the (Newtonian-limit normalized) spin-2 vertex contraction
 factor, $2 \, u_1^{\alpha} u_1^{\beta} P_{\alpha \beta ;\alpha' \beta'} u_2^{\alpha'} u_2^{\beta'}$, so that 
\be\label{chispin0}
\frac12 \chi_0 = \frac{ G_0 m_1 m_2}{ J } \, \frac1{\widehat {\mathcal D}}\,.
\ee
Here $G_2$ denotes the spin-2 contribution to the Cavendish constant and $G_0$ its spin-0 contribution. 
Let us parametrize (in keeping with the notation of Ref. \cite{Damour:1992we}) the admixtion of scalar exchange to the gravitational interaction by the fraction
\be
\alpha^2 \equiv \frac{G_0}{G_2}\,.
\ee
This notation means that each scalar-matter vertex carries an extra factor $\alpha$, with respect to a tensor-matter one. 

The total observable Cavendish constant then reads
\be
G= G_2 + G_0 = (1 + \alpha^2) \, G_2 \,,
\ee
while the total scattering function $\chi=\chi_2 + \chi_0$  reads [using Eqs. \eqref{D2}, \eqref{chireal2}, \eqref{epsilon};
i.e. $\epsilon(s) = - u_1.u_2$]
\bea
\frac12 \chi(s, J)&=&  \frac{  m_1 m_2}{ J  \, {\widehat {\mathcal D}}} \left[ G_2 \left( 2 (u_1.u_2)^2 -1 \right) + G_0 \right] \nonumber\\
&=& \frac{ G m_1 m_2}{ J }  \frac{ (1+ \gamma) \, \epsilon^2(s) - \gamma }{ \sqrt{\epsilon^2(s) - 1} }\,,
\eea
where  $1+\gamma = 2 G_2/G= 2/(1+\alpha^2)$, i.e.
\be \label{gamma}
\gamma = \frac{1 - \alpha^2}{1+ \alpha^2}\,.
\ee
Comparing with Eq. \eqref{chieff3}, and going through the reasoning used in the previous section, we can then conclude that, to first post-Minkowskian order, the
two-body interaction in a theory of
gravity combining (massless) tensor and scalar fields can be described by the following modifications of the usual EOB theory:
(i) the energy map $f$ is (to all orders in $v/c$) again given by the  result \eqref{fx}; while, 
(ii) the (linearized) effective metric differs from the (linearized) Schwarzschild metric by the presence of a coefficient $\beta_1=\gamma$, Eq. \eqref{gamma},
in the spatial metric, i.e.
\bea \label{geffgamma}
g_{\mu \nu}^{\rm eff}dx^{\mu} dx^{\nu} &=& -\left(1- \frac{2 G M}{R}\right) \,dt^2 + \left(1+ \gamma \frac{2 G M}{R}\right) \,dR^2 \nonumber\\
&+& R^2(d\theta^2 + \sin^2\theta \, d \varphi^2) + O(G^2)\,.
\eea
This modification of the linearized Schwarzschild metric is equivalent [modulo the first-order coordinate change $R= R' + \gamma GM + O(G^2)$] to the $G$-linearization of the usual  1PN-accurate Eddington, parametrized-post-Newtonian metric
\bea
ds_{\rm PPN}^2 &=& - \left(1 - 2 \frac{ G M}{R'} + 2 \beta \left(\frac{ G M}{R'}\right)^2 \right) dt^2  \nonumber\\
 &+&  \left(1 + 2 \gamma \frac{ G M}{R'} \right) \left(dR'^2+ R'^2(d\theta^2 + \sin^2\theta \, d \varphi^2) \right) \,. \nonumber\\ 
\eea
The result \eqref{gamma} then agrees with the expression of the Eddington parameter $\gamma$ in terms of the linear scalar coupling
$\alpha$, often also written as [see, e.g., Eq. (4.12c) in Ref. \cite{Damour:1992we}]
\be
\bar \gamma \equiv \gamma -1 = - 2 \frac{ \alpha^2}{1+ \alpha^2} \,.
\ee
To see the $O(G^2)$ Eddington parameter $\beta \equiv 1 + \bar \beta $ one would have to introduce a nonlinear vertex $ -\frac12  { \beta_0} \int m_a \phi^2(x_a) ds_a$ coupling the scalar field $\phi(x)$ to the worldlines ($a=1,2$), and to go beyond the $G$-linear approximation.
[In the 1PN approximation, this yields the link $ \bar \beta = +\frac12 \beta_0  \alpha^2/(1+ \alpha^2)^2$  \cite{Damour:1995kt}.]

\section{Conclusions}

A new approach to the effective one-body description of gravitationally interacting two-body systems has been introduced.
This approach is not based, as previous work,  on the post-Newtonian approximation scheme (combined expansion in $\frac{G}{c^2}$ 
and in $\frac{1}{c^2}$), but rather on the {\it post-Minkowskian} approximation scheme 
(perturbation theory in $G$, without assuming small velocities). It uses a new dictionary based on considering the functional
dependence of the scattering angle $\chi$ on the total energy and total angular momentum of the system (all quantities
being considered in the center of mass frame). By explicitly calculating (in a novel way, analogous to quantum-scattering-amplitude computations)
the  first post-Minkowskian [$O(G)$] scattering function, we have proven {\it to all orders in $v/c$}  two results that were previously only known to
a limited post-Newtonian accuracy: (i) To order $G^1$, the relativistic dynamics of a two-body system (of masses $m_1, m_2$)
is equivalent to the relativistic dynamics of an effective test particle of mass $\mu= m_1 m_2/(m_1+m_2)$ moving in a {\it Schwarzschild metric
of mass $M=m_1 +m_2$}; and (ii) This equivalence requires the existence of an {\it exactly quadratic map} $f$ between the real 
(relativistic) two-body energy ${\mathcal E}_{\rm real}$ and the (relativistic) energy ${\mathcal E}_{\rm eff}$ of the effective particle
given (to all orders in $1/c$) by
\be \label{fEOB2}
{\mathcal E}_{\rm eff}= \frac{({\mathcal E}_{\rm real})^2 - m_1^2 c^4 -m_2^2 c^4}{2 \, (m_1+m_2) c^2} \, .
\ee
The latter energy map was also proven to apply to the effective one-body description of two masses interacting in tensor-scalar
gravity [for which one must use an $O(G)$ effective metric modified by Eddington's $\gamma$ parameter].

We leave to future work the generalization of our approach to higher orders in $G$, and to spinning bodies.

\section*{Acknowledgments}
I am grateful to Ulysse Schildge for his help during many years.

\end{document}